\documentclass[prl,aps,showpacs,twocolumn,floatfix]{revtex4}
\usepackage{epsfig} \usepackage{graphics} \usepackage{bm}
\usepackage{amssymb}
\usepackage{graphicx}
\begin{document}

\preprint{Lebed-PRL-LN}

\title{Unconventional Field-Induced Spin-Density-Wave Phases in
Quasi-One-Dimensional Conductors in High Magnetic Fields}

\author{A.G. Lebed$^*$}

\affiliation{Department of Physics, University of Arizona, 1118 E.
4-th Street, Tucson, AZ 85721, USA}

\begin{abstract}
It is known that the Field-Induced Spin-Density-Wave (FISDW)
phases are experimentally observed in the quasi-one-dimensional
(Q1D) organic conductors with chemical formula (TMTSF)$_2$X
(X=PF$_6$, ClO$_4$, etc.) and some others in moderate magnetic
fields. From a theoretical point of view, they appear as a result
of "one-dimensionalization" of the Q1D electron spectra due to the
orbital electron effect in a magnetic field. We predict that the
novel FISDW phases with different physical meaning have to appear
in inclined high magnetic fields in Q1D conductors as a result of
combination of the spin-splitting and orbital electron effects. We
suggest performing the corresponding experiments in the
(TMTSF)$_2$X materials.

\end{abstract}


\maketitle

We recall that the so-called Field-Induced Spin-Density-Wave
(FISDW) phases and related 3D quantum Hall effect (3D QHE) are
experimentally observed in a number of Q1D organic conductors
[1,2]. The most studied among them are compounds with chemical
formula (TMTSF)$_2$X (X=PF$_6$, ClO$_4$, etc.), where the above
mentioned phenomena were first discovered [3,4]. Theory of the
FISDW phases was successfully developed in Refs.[5-15], whereas
the 3D-QHE was considered in Refs.[13,14].

It is known [7,1,2,5,6] that electron spectrum of the Q1D organic
conductors (TMTSF)$_2$X can be well described by the following
tight-binding approximation:
\begin{eqnarray}
\epsilon^{\pm}({\bf p}) = \pm v_F( p_x \mp p_F) -2t_b \cos(p_y
b^*)
\nonumber\\
-2t'_b \cos(2p_y b^*) -2t_c \cos(p_z c^*).
\end{eqnarray}
[Here, $p_F$ and $v_F$ are the electron Fermi momentum and Fermi
velocity along the conducting axis ${\bf a^*}$, $t_b$ and $t_c$
are the electron wave functions overlapping integrals along axes
${\bf b^*}$ and ${\bf c^*}$, respectively, and the term with
$t'_b$ appears due to non-linearity of the real Q1D electron
spectrum along ${\bf a}$ axis [7,6] ($p_F v_F \gg t_b \gg t'_b,
t_c$)]. It is important that, for $t'_b=0$, the electron spectrum
(1) possesses the so-called "nesting" property [7,1,2,5,6],
\begin{equation}
\epsilon_{\sigma}^{+}(\Delta p_x,p_y,p_z) +
\epsilon_{-\sigma}^{-}(\Delta p_x,p_y+\pi/b^*,p_z+\pi/c^*) = 0 ,
\end{equation}
which allows to stabilize Spin-Density-Wave (SDW) phase due to the
Peierls instability with wave vector [7,2,5,6]:
\begin{equation}
Q_0 = (2p_F,\pi/b^*,\pi/c^*),
\end{equation}
where $\sigma=+(-)$ for electron spin up(down), respectively.

We stress that the so-called "anti-nesting" term, $2t'_b \cos(2p_y
b^*)$, in Eq.(1) destroys [7,1,2,5,6] the "nesting" condition (2)
and, therefore, for high enough values of the parameter $t'_b$,
metallic or superconducting phases can become the ground states.
In this case, as experimentally shown in Refs.[3,4], the physical
situation is very interesting in a magnetic field applied along
axis ${\bf c^*}$: a cascade of the numerous FISDW phases occurs
with Hall conductivity being quantized. In theoretical works
[5-15], it was shown that the "one-dimensionalized" [5] orbital
electron motion in the magnetic field restores the Peierls
instability and these FISDW phases were interpreted as the SDW
phases with the following quantized wave vectors:
\begin{equation}
Q_n = (2p_F + n \ \omega_c/v_F,\pi/b^*,\pi/c^*),
\end{equation}
where $n$ is an integer.

Note that, using the linearized Q1D electron spectrum (1), it is
possible to come to the conclusion [5,6,8-15] that the Pauli
spin-splitting effect plays no role in physical properties of the
FISDW phases. Very recently, we have shown [16] that, due to
non-linearity of the electron spectrum along the conducting axis
${\bf a^*}$, the Pauli spin-splitting effect generates a new
"anti-nesting" term in the Q1D spectrum (1). In particular, in
Ref.[16], we have demonstrated that this term results in a
destruction of the SDW phase in a parallel magnetic field, ${\bf
H} \parallel {\bf a^*}$ (i.e., in the absence of the orbital
effect).

 The goal of our Rapid Communication is to show that the above mentioned
new "anti-nesting" term [16], which is proportional to a strength
of a magnetic field, is responsible for even more interesting
phenomenon - a novel cascade of the FISDW phases in high magnetic
fields. Unlike the known FISDW phases, the new ones have to appear
due to simultaneous actions of the Pauli spin-splitting and
orbital electron effects. We suggest to observe the novel FISDW
phases in the organic conductors from chemical family (TMTSF)$_2$X
in an inclined with respect to the conducting axis ${\bf a^*}$
magnetic field. In particular, we show that, unlike the known
cascade of the FISDW phases [1-15], where the FISDW phase
boundaries are roughly periodic with an inverse magnetic field,
positions of the novel FISDW phases do not depend on a strength of
the magnetic field. Below, they are shown to depend almost
periodically on an inverse sinusoidal function of the inclination
angle, $1/\sin \alpha$.

For further development, we consider the following 2D
non-linearized model of Q1D spectrum in a parallel to the
conducting ${\bf a^*}$ axis magnetic field,
\begin{equation}
\epsilon_{\sigma}({\bf p}) = -2t_a \cos(p_xa^*/2)-2t_b \cos(p_y
b^*) - \mu_B \sigma H ,
\end{equation}
where $t_a$ is overlapping integral of electron wave functions
along the conducting ${\bf a^*}$ axis. [Note that the first
"anti-nesting" term, $2 t'_b \cos(2p_yb^*)$, appears later as a
result of a non-linearity of the spectrum (5) with respect to
variable $p_x$.] Contrary to the all existing theories of the
FISDW phases [5,6,8-15], below we take account of both linear and
quadratic terms of the variables $(p_x \pm p_F)$ near two pieces
of the Fermi surface in a parallel magnetic field:
\begin{equation}
\epsilon^{\pm}(p_x) = \pm v_F (p_x \mp p_F) + \gamma (p_x \mp
p_F)^2 ,
\end{equation}
where in the (TMTSF)$_2$X compounds [1,2,7,16]
\begin{equation}
v_F = \frac{t_a a^*}{\sqrt{2}} , \ \ \ p_F = \frac{\pi}{2a^*}, \ \
\ \gamma = \frac{t_a (a^*)^2}{4 \sqrt{2}} .
\end{equation}
Using Eqs.(5)-(7), it is easy to demonstrate that, in a parallel
magnetic field, near two sheets of the FS the electron spectrum
can be written as
\begin{eqnarray}
\epsilon^{\pm}_{\sigma}({\bf p}) = \pm v_F(p_x \mp p_F) +
t^{\pm}_{b}(p_y,\sigma) -\mu_B \sigma H + \Delta \epsilon ,
\nonumber\\
t^{\pm}_b(p_y, \sigma) = - 2t_b \cos(p_y b^*) + 2 t'_b \cos(2 p_y
b^*)
\nonumber\\
+ 2t_H \sigma \cos(p_y b^*),
\end{eqnarray}
where $ t_H = \mu_B H t_b/(\sqrt{2} t_a), \ t'_b =  t^2_b/(2
\sqrt{2}t_a), \ \Delta \epsilon = \mu^2_B H^2/(2 \sqrt{2} t_a)$.

\begin{figure}[t]
\centering
\includegraphics[width=0.5\textwidth]{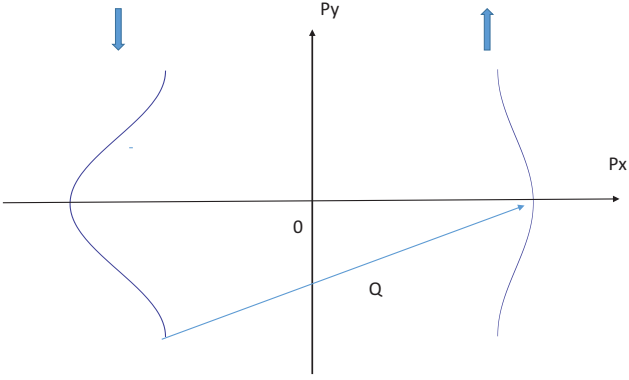}
\caption{Left and right parts of the Q1D Fermi surface, given by
Eq.(8) at $t'_b=0$, are shown for different directions of electron
spins (i.e., for the SDW paring). As seen, that even in this case,
where the traditional "anti-nesting" term disappears, there
appears the second "anti-nesting" term, which destroys the
"nesting" properties of the Q1D spectrum for any SDW wave-vector
${\bf Q}$.}
\end{figure}

It is important that Eq.(8) contains two kinds of "anti-nesting"
terms for the SDW instability (pay attention that the second term
does not exist for a CDW case and, thus, does not destroy the CDW
phase). The first of them is usual, $2 t'_b \cos(2 p_y b^*)$,
which is responsible for the standard FISDW cascade of phase
transitions and 3D QHE [5-15], provided the magnetic field has a
finite ${\bf c^*}$ axis component The second "anti-nesting" term,
$2t_H \sigma \cos(p_y b^*)$ (see Fig.1), has magnetic field
dependent amplitude, $t_H \sim H$, which as we show generates the
novel FISDW phases in high magnetic fields. In contrast, terms
$-\mu_BH$ and $\Delta \epsilon$ in Eq.(8) do not affect seriously
the FISDW phases. Indeed, term $-\mu_BH$ disappears for the SDW
pairing, whereas term $\Delta \epsilon$ just shifts the wave
vectors of the FISDW phases. Therefore, we omit the last two terms
in our further calculations.

We stress that the geometry suggested by us for observations of
the novel FISDW phases is different from the standard one
[5,6,8-15] as well as different from that in Ref.[16]. In our
case, we have strong magnetic field, which is characterized by
both finite ${\bf a^*}$ and ${\bf c^*}$ axes components:
\begin{equation}
{\bf H} = (H \cos \alpha, 0, H \sin \alpha),
\end{equation}
\begin{equation}
{\bf A} = (0, H x \sin \alpha , H y \cos \alpha),
\end{equation}
where $\alpha$ is angle between magnetic field ${\bf H}$ and ${\bf
a^*}$ axis. Under such experimental conditions, we have both the
Pauli spin-splitting effect (8) and the orbital one [5-15]. Below,
we introduce the orbital effect by means of the so-called Peierls
substitution method, as it is done in Ref.[5]:
\begin{eqnarray}
p_x \mp p_F \rightarrow -i \frac{d}{dx}, \ \ \ p_yb^* \rightarrow
p_yb^* - \frac{e}{c} A_yb^*
\nonumber\\
= p_yb^* - \frac{\omega_c(\alpha)x}{v_F}, \ \ \
\omega_c(\alpha)=\frac{ev_Fb^*H\sin\alpha}{c}.
\end{eqnarray}

Let us introduce slow varying parts, $g^{\pm \pm}(i \omega_n, p_y;
x, x_1; \sigma)$, of the non-interacting electron Green's
functions near two open sheets of the Q1D FS, $G^{\pm \pm}(i
\omega_n, p_y; x, x_1; \sigma)$, by the following equations:
\begin{eqnarray}
G^{++}(i \omega_n, p_y; x, x_1; \sigma) = e^{ip_F(x-x_1)}
\nonumber\\
\times g^{++}(i \omega_n, p_y; x, x_1; \sigma),
 \end{eqnarray}
\begin{eqnarray}
 G^{--}(i \omega_n, p_y; x, x_1; \sigma) = e^{-ip_F(x-x_1)}
 \nonumber\\
\times g^{--}(i \omega_n, p_y; x, x_1; \sigma),
\end{eqnarray}
where $\omega_n$ is the Matsubara's frequency [17]. Then, using
Eqs.(8) and (11), it is possible to make sure that the slow
varying parts of the electron Green's functions obey the
equations:
\begin{eqnarray}
\biggl\{ i \omega_n +i v_F \frac{d}{dx} -t^+_b \biggl[ p_yb^* -
\frac{\omega(\alpha)x}{v_F}, \sigma \biggl] \biggl\}
\nonumber\\
\times g^{++}(i \omega_n, p_y; x, x_1; \sigma) = \delta (x-x_1) \
,
\end{eqnarray}
\begin{eqnarray}
\biggl\{ i \omega_n - i v_F \frac{d}{dx} -t^-_b \biggl[ p_yb^* -
\frac{\omega(\alpha)x}{v_F}, \sigma \biggl] \biggl\}
\nonumber\\
\times g^{--}(i \omega_n, p_y; x, x_1; \sigma) = \delta (x-x_1) \
,
\end{eqnarray}
where $\delta(x-x_1)$ is the Dirac's delta-function. It is
important that the equations for slow varying parts of the Green's
functions of non-interacting electrons in a magnetic field (14)
and (15) can be exactly solved:
\begin{eqnarray}
g^{++}(i \omega_n, p_y; x, x_1; \sigma)= \frac{sgn (\omega_n)}{i
v_F} \exp \biggl\{ -\frac{\omega_n(x-x_1)}{v_F} \nonumber\\
-\frac{i}{v_F} \int^x_{x_1} t^+_b \biggl[ p_y b^* -
\frac{\omega(\alpha)u}{v_F} , \sigma \biggl]  du \biggl\} , \
\omega_n (x-x_1)
> 0,
\end{eqnarray}
\begin{eqnarray}
g^{--}(i \omega_n, p_y; x, x_1; \sigma)= \frac{sgn (\omega_n)}{i
v_F} \exp \biggl\{ \frac{\omega_n(x-x_1)}{v_F}
\nonumber\\
+\frac{i}{v_F} \int^x_{x_1} t^-_b \biggl[ p_y b^* -
\frac{\omega(\alpha)u}{v_F} , \sigma \biggl]  du \biggl\} , \
\omega_n (x-x_1) < 0.
\end{eqnarray}

Here, we calculate a linear response of electrons to the external
field, corresponding to the following SDW electron-hole pairing,
\begin{equation}
\hat h ({\bf Q}) = (\hat \sigma_x)_{\alpha \beta} \ \exp(i {\bf Q}
{\bf r}) \ ,
\end{equation}
in a similar way as it is done in Ref.[5] for different Q1D
spectrum without the magnetic field dependent term. In random
phase approximation, we obtain the so-called Stoner equation for
susceptibility:
\begin{equation}
\chi ({\bf Q}) = \frac{\chi_0 ({\bf Q})}{[1 - g \chi_0 ({\bf
Q})]}.
\end{equation}
In Eq.(19), $g$ is the effective SDW electron coupling constant,
$\chi_0 ({\bf Q})$ is a susceptibility of the non-interacting
electrons in a magnetic field:
\begin{eqnarray}
&&\chi_0(\approx 2p_F, Q_y, \pi/c^*) = T \sum_{\omega_m}
\sum_{\sigma} \int \frac{dp_y}{2 \pi} \int dx_1
\nonumber\\
&&\times g^{++}(i \omega_n, p_y; x, x_1; \sigma) \ g^{--}(i
\omega_n, p_y -Q_y; x_1, x; -\sigma).
\end{eqnarray}

Let us substitute the known slow varying parts of the electron
Green's functions in a magnetic field (16) and (17) in Eqs.(20)
and (19). As a result of straightforward but lengthy calculations,
we obtain the following equation of a stability for the FISDW
phases in the presence of the orbital and Pauli spin-splitting
effects:
\begin{eqnarray}
&&\frac{1}{g}= \max_{ (k, \Delta t)}
\int^{\infty}_{\frac{v_F}{\Omega}} \frac{2 \pi T_c dz}{v_F \sinh
\biggl( \frac{2 \pi T_c z}{v_F} \biggl)} \biggl< \cos \biggl\{
\biggl[ \frac{8 \Delta t}{\omega_c(\alpha)}\biggl] \sin
\biggl[\frac{\omega_c(\alpha)z}{2v_F} \biggl]
\nonumber\\
&&\times \sin(p_y b^*) - \biggl[ \frac{4 t'_b}{\omega_c(\alpha)}
\biggl] \sin \biggl[\frac{\omega_c(\alpha)z}{v_F} \biggl]
\cos(2p_yb^*)  - k z \biggl\}
\nonumber\\
&&\times \cos \biggl\{ \biggl[\frac{8t_H}{\omega_c(\alpha)}
\biggl] \sin \biggl[\frac{\omega_c(\alpha)z}{2v_F} \bigg] \cos(p_y
b^*) \biggl\} \biggl>_{p_y} .
\end{eqnarray}
[Here, $Q_x = 2p_F +k$, $Q_y = \pi/b^* + q \ (qb^* \ll 1)$,
$\Delta t = t_b q b^*/2$, and $\Omega$ is a cutoff energy;
$<...>_{p_y}$ stands for averaging procedure over variable $p_y$].
We stress that, in Eq.(21), we maximize SDW transition
temperature, $T_c$, with respect to longitudinal, $k$, and
transverse, $q$, wave vectors under the condition that $t_b \gg
t'_b$.

It is important that Eq.(21), derived in our Rapid Communication,
is the most general equation to determine the appearance of the
FISDW phases in a Q1D conductor in a magnetic field. In
particular, at low enough magnetic fields, we can disregard the
Pauli spin-splitting effect and, therefore, at $t_H=0$ and
$\alpha= \pi/2$, Eq.(21) coincides with the main equation of
Ref.[10]. Our current goal is not a full analysis of Eq.(21),
which is difficult numerical problem and hopefully will be solved
in the future. In the Rapid Communication, we consider high
magnetic field limit of Eq.(21) to demonstrate novel phenomenon -
the appearance of high magnetic field FISDW phases due to the the
combination of the orbital electron motion and Pauli
spin-splitting effects. To this end, let us consider high magnetic
fields, where
\begin{equation}
\omega_c(\alpha) \geq 4t'_b.
\end{equation}
It is easy to see that, in this limit, the master Eq.(21) can be
rewritten as
\begin{eqnarray}
&&\frac{1}{g}= \max_{ (k, \Delta t)}
\int^{\infty}_{\frac{v_F}{\Omega}} \frac{2 \pi T_c dz}{v_F \sinh
\biggl( \frac{2 \pi T_c z}{v_F} \biggl)} \biggl< \cos \biggl\{
\biggl[ \frac{8 \Delta t}{\omega_c(\alpha)}\biggl]
\nonumber\\
&&\times \sin \biggl[\frac{\omega_c(\alpha)z}{2v_F} \biggl]
\sin(p_y b^*)  - k z \biggl\}
\nonumber\\
&&\times \cos \biggl\{ \biggl[\frac{8t_H}{\omega_c(\alpha)}
\biggl] \sin \biggl[\frac{\omega_c(\alpha)z}{2v_F} \bigg] \cos(p_y
b^*) \biggl\} \biggl>_{p_y} .
\end{eqnarray}

Let us simplify the integral (23), using some well known
trigonometric equations. As a result of rather simple
calculations, instead of Eq.(23), we obtain
\begin{eqnarray}
&&\frac{1}{g}= \max_{ (k, \Delta t)}
\int^{\infty}_{\frac{v_F}{\Omega}} \frac{2 \pi T_c dz}{v_F \sinh
\biggl( \frac{2 \pi T_c z}{v_F} \biggl)} \ \cos(kz)
\nonumber\\
&&\times J_0 \biggl\{ \sqrt{\biggl[ \frac{8 \Delta
t}{\omega_c(\alpha)} \biggl]^2 + \biggl[ \frac{8
t_H}{\omega_c(\alpha)} \biggl]^2} \times \sin
\biggl[\frac{\omega_c(\alpha) z}{2v_F} \biggl] \biggl\}.
\end{eqnarray}
Note that in the derivation of Eq.(24) from Eq.(23) we also take
into account that [18]:
\begin{equation}
\int^{+ \pi}_{-\pi}\frac{d \phi}{2 \pi} \exp[iA
\sin(\phi)]=J_0(A).
\end{equation}
From Eq.(24), it is evident that the integral takes the maximal
value at $\Delta t=0$ (i.e., at $Q_y=\pi/b^*$). In this case, we
can rewrite stability condition for the FISDW phases (24) as
\begin{eqnarray}
&&\frac{1}{g}= \max_{k} \int^{\infty}_{\frac{v_F}{\Omega}} \frac{2
\pi T_c dz}{v_F \sinh \biggl( \frac{2 \pi T_c z}{v_F} \biggl)} \
\cos(kz)
\nonumber\\
&&\times J_0 \biggl\{ \biggl[ \frac{8 t_H}{\omega_c(\alpha)}
\biggl] \ \sin \biggl[\frac{\omega_c(\alpha) z}{2v_F} \biggl]
\biggl\},
\end{eqnarray}
where $J_0(x)$ is the Bessel function of the zeroth-order.

Below, we use one more relationship between the trigonometric and
Bessel functions of the $n$-order, $J_n(x)$, [18]:
\begin{equation}
J_0[2x\sin(\phi)] = \sum^{\infty}_{n=-\infty} J^2_n(x)
\cos(2n\phi),
\end{equation}
where $J_{-n}(x)=(-1)^n J_n(x)$. Using comparison of Eqs.(26) and
(27), it is easy to come to the following important conclusion.
The integral in Eq.(26) possesses logarithmic divergencies at low
temperatures for the following quantized values of the
longitudinal wave vector:
\begin{equation}
Q_x = 2p_F + k, \ \ \ k= n \ \omega_c(\alpha)/v_F,
\end{equation}
where $n$ is an arbitrary integer number. This statement has two
consequences. First consequence is that one of the FISDW phases
(28) is always a ground state of our system at $T=0$ unless
$\alpha =0$ in Eq.(9), which validates the main statement of the
Rapid Communication about the appearance of a novel cascade of the
FISDW phases in high magnetic fields. Second consequence is that,
at low but finite temperatures, the wave vector, which results in
maximum of the FISDW transition temperature, is close to (28).
Therefore, in high magnetic fields (i.e., low temperatures), where
\begin{equation}
\omega_c(\alpha) \gg T_c ,
\end{equation}
we can rewrite Eq.(26) as
\begin{eqnarray}
&&\frac{1}{g}= \max_{n} \int^{\infty}_{\frac{v_F}{\Omega}} \frac{2
\pi T_c dz}{v_F \sinh \biggl( \frac{2 \pi T_c z}{v_F} \biggl)} \
\cos[n \omega_c(\alpha) z /v_F]
\nonumber\\
&&\times J_0 \biggl\{ \biggl[ \frac{8 t_H}{\omega_c(\alpha)}
\biggl] \ \sin \biggl[\frac{\omega_c(\alpha) z}{2v_F} \biggl]
\biggl\},
\end{eqnarray}
where we have to find maximum of the integral (30) with respect to
the integer $n$ in Eq.(28).

Let us consider below some interesting limiting case, which has a
clear physical meaning. Suppose that we are interested in case,
where magnetic field is characterized by small inclinations from
conducting ${\bf a^*}$ axis [i.e., the case of small $\alpha$ in
Eq.(9)]. Then, in the limit
\begin{equation}
4t_H \geq \omega_c(\alpha)
\end{equation}
it possible to make sure that Eq.(30) can be rewritten with
logarithmic accuracy as
\begin{eqnarray}
\frac{1}{g}= \max_{n} \biggl\{ \int^{\infty}_{\frac{v_F}{\Omega}}
\frac{dz}{z} J_0 \biggl(\frac{4t_Hz}{v_F} \biggl)\cos
\biggl[\frac{n \omega_c(\alpha)}{v_F} \bigg]
\nonumber\\
+ J^2_n \biggl[ \frac{4 t_H}{\omega_c(\alpha)} \bigg] \ln
\biggl[\frac{\omega_c(\alpha)}{T_c} \biggl]\biggl\}.
\end{eqnarray}
Note that two different terms in Eq.(32) have completely different
physical meanings. Indeed, the first term was obtained before in
Ref.[16] and describes destruction of SDW phase by a magnetic
field due the appearance of the so-called magnetic field dependent
"anti-nesting" term, $t_H$. On the other hand, the second term
describes the above discussed logarithmic divergencies of the
integral (26) for the quantized FISDW wave vectors (28). As we
discussed above, the second term makes the quantized FISDW phases
to be ground states at low enough temperatures and high enough
magnetic fields. Therefore, contrary to the main conclusion of
Ref.[16], which is valid only for $\alpha =0$ (i.e., in a parallel
magnetic field), in our case, almost destroyed SDW phase restores
as a cascade of the FISDW phases in high enough magnetic fields.

The first term in the integral (32) was considered in details in
Ref.[16]. In particular, it was shown that it takes maximum for
all wave vectors from the following interval,
\begin{equation}
\frac{n \ \omega_c(\alpha)}{v_F} < \frac{4 t_H}{ v_F},
\end{equation}
and this maximum is equal to $\ln (H_0/H)$, where $H_0$ is the
critical magnetic field, which destroys SDW phase at $\alpha=0$.
Fortunately, it is possible to make sure that the second term
takes maximum for some integers numbers $n$ from the interval
(33). Therefore, Eq.(32) can be rewritten as
\begin{equation}
\ln \biggl( \frac{H}{H_0} \biggl) = \max_n \biggl\{ J^2_n
\biggl[\frac{4t_H}{\omega_c(\alpha)} \biggl] \bigg\} \ln \biggl[
\frac{\omega_c(\alpha)}{T_c} \biggl], \ H > H_0.
\end{equation}

\begin{figure}[t]
\centering
\includegraphics[width=0.5\textwidth]{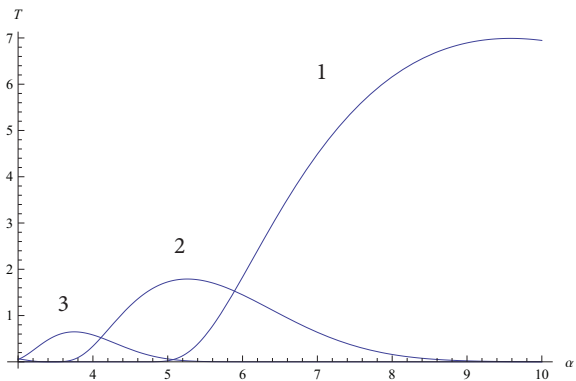}
\caption{Transition temperature to the unconventional FISDW phases
(measured in Kelvin) is calculated by means of Eq.(35) as a
function of angle $\alpha$ (measured in degrees) and is given by a
maximum value of three curves. Numbers 1,2,3 correspond to the
values of integer $n$ for the quantized FISDW wave vector in
Eq.(28). The calculations are made for $H = 350 \ T$ and the
following realistic values of the physical parameters in the
(TMTSF)$_2$PF$_6$: $\omega_c(\pi/2)/H = 0.85 \ K/T$ and $H_0
\simeq 185 \ T$.}
\end{figure}

Then, from Eq.(34), it directly follows that transition
temperature to the different FISDW phases (28) with the
logarithmic accuracy can be expressed as
\begin{equation}
T_c = \omega_c(\alpha) \ \exp \biggl( - \frac{\ln (H/H_0)}{ \max_n
\biggl\{
 J^2_n [4t_H/\omega_c(\alpha)] \biggl\}} \bigg), \ H > H_0,
\end{equation}
where we have to take maximum of $T_c$ with respect to the integer
$n$ to find a ground FISDW state (see Fig.2). As directly follows
from Eq.(35), the physical meaning of the novel FISDW phases is
different from that of standard ones [5-15]. First of all, in
Eq.(35), we have unusual "anti-nesting" term and related parameter
$t_H \sim H$. Since $\omega_c(\alpha)$ is also proportional to
$H$, the positions of the different FISDW phases, given by
Eq.(35), do not depend on a strength of a magnetic field, in
contrast to the standard case [1-6,7-15]. In addition, the
following asymptotic property of the Bessel functions for $z \gg
1$ [18],
\begin{equation}
J_n(z) \approx \sqrt{\frac{2}{\pi z}} \cos \biggl(z
-\frac{n\pi}{2} - \frac{\pi}{4} \biggl)
\end{equation}
shows that the positions of the different FISDW phases at large
values of integer $n$ in Eq.(28) are periodic with variable
$1/\sin \alpha$.

In conclusion, let us discuss the possible applications of our
theory to real Q1D conductors from the chemical family
(TMTSF)$_2$X. We recall that, in Ref.[16], we have suggested novel
effect, where SDW state is destroyed by some unusual
"anti-nesting" term. It appears due to the Pauli spin-splitting
effects [see the second "anti-nesting" term in Eq.(8)]. The theory
[16] was elaborated for a magnetic field parallel to the
conducting axis, ${\bf H}
\parallel {\bf a}$ (i.e., for $\alpha =0$ in our case). As
directly seen from Eq.(21), for any $H \neq 0$ and $\alpha \neq 0$
the FISDW phases are ground states at $T=0$. Therefore, for the
analysis of the FISDW phases we always have to make use of our
Eq.(21), which is more general than that used in the previous
theories of the FISDW phases [5,6,8-15]. Below, we discuss where
our final analytical Eq.(35) is literally applicable to describe
novel FISDW phases suggested in the Letter. As an example, let us
consider (TMTSF)$_2$PF$_6$ conductor. First, the magnetic field
has to be stronger than the value of $H_0$, which is estimated as
$H_0 =185 \ T$ (see Ref.[16]). Note that such high magnetic fields
are now available (see, for example, Refs.[19,20]). Second, angle
$\alpha$ has to be not very small in order inequality (22) to be
fulfilled. If we estimate the value of $t'_b \approx 5 \ K$ from
the expression $t'_b=t^2_b/(2\sqrt{2} t_a)$, then we obtain from
Eq.(22) that $\alpha \geq 6^0$. It is possible to make sure that
the inequality (31) is also fulfilled under the above mentioned
conditions. Note that above we have estimated the values of a
magnetic field, where the novel FISDW phases will appear,
nevertheless our Eq.(21) predicts some changes in the known
cascades of the FISDW phases. Hopefully, they will be studied by
using numerical methods in the future.

 We are  thankful to N.N. Bagmet (Lebed) for useful discussions.

$^*$Also at: L.D. Landau Institute for Theoretical Physics, RAS, 2
Kosygina Street, Moscow 117334, Russia.

\end{document}